\documentclass[sigconf, nonacm]{acmart}

\newcommand\vldbdoi{XX.XX/XXX.XX}

\newcommand\vldbpagestyle{plain}

\usepackage{graphicx}
\usepackage{subcaption}
\usepackage{listings}
\usepackage{enumitem}
\usepackage{booktabs}
\usepackage{pifont}
\usepackage{makecell}
\usepackage[most]{tcolorbox}
\usepackage{microtype}

\newcommand{\minihead}[1]{{\vspace{.55em}\noindent\textbf{#1.} }}
\newcommand{\bird}{\textsc{Bird }} 

\newcommand{\mycross}{\textcolor{red}{\textbf{$\times$}}}
\newcommand{\mycheck}{\textcolor{ForestGreen}{\textbf{\checkmark}}}
\newcommand{\claude}{Claude-3.5-Sonnet }
\newcommand{\syntaxhighlight}[1]{\textbf{\textcolor{blue}{#1}}}
\newcolumntype{C}[1]{>{\centering\arraybackslash}m{#1}}

\newcommand{\LeftCentered}[1]{%
  \makebox[2cm][c]{\makebox[0.8cm][l]{#1}}%
}

\newcommand{\LeftCenteredx}[1]{%
  \makebox[2cm][c]{\makebox[0.7cm][l]{#1}}%
}

\newcommand{\LeftCenteredxx}[1]{%
  \makebox[2cm][c]{\makebox[0.8cm][l]{#1}}%
}

\newcommand{\LeftCenteredxxx}[1]{%
  \makebox[2cm][c]{\makebox[0.5cm][l]{#1}}%
}

\newcommand*\LSTfont{\Small\ttfamily\SetTracking{encoding=*}{-60}\lsstyle}

\lstdefinestyle{yaml}{
    frame=single,
    language=,
    basicstyle=\LSTfont, 
    keywordstyle={[2]\color{blue}\bfseries}, 
    sensitive=true,
    literate=
        {name:}{{\color{blue}\bfseries name:}}1
        {description:}{{\color{blue}\bfseries description:}}1
        {columns:}{{\color{blue}\bfseries columns:}}1
        {models:}{{\color{blue}\bfseries models:}}1
}

\lstdefinestyle{yaml}{
    frame=single,
    language=,
    basicstyle=\LSTfont,
    morekeywords={[2]{name, description, columns, models}},
    keywordstyle={[2]\color{blue}},
    sensitive=true,
    literate=
      {:}{{{\color{blue}{:}}}}1
}

\lstdefinestyle{yamll}{
    frame=single,
    language=,
    basicstyle=\LSTfont,
    sensitive=true,
    morekeywords={[3]Airbyte, config, files_definition_id, workspace_id, flat_files, format, path, sync_mode, table, snowflake, account, database, password, role, schema, username, warehouse, connection_string, mongodb}, 
    morekeywords={[2]'mongodb,elt-mongodb}, 
    keywordstyle={[3]\color{blue}},   
    keywordstyle={[2]\color{black}},  
    alsoletter={'-}
}

\lstdefinestyle{terraform}{
    frame=single,
    language=,
    basicstyle=\LSTfont,
    morekeywords={[2]{resource, name, definition_id , workspace_id, configuration, provider, https_public_web, url, format, dataset_name, source_id, destination_id, namespace_definition, configurations, streams, sync_mode, database_config, self_managed_replica_set, connection_string }},
    morekeywords={[3].destination_id,.source_id}, 
    keywordstyle={[2]\color{blue}},
    keywordstyle={[3]\color{black}},
    sensitive=true,
    alsoletter={.}
}

\lstdefinestyle{dbt}{
    frame=single,
    language=,
    basicstyle=\LSTfont,
    morekeywords={[2]{my_dbt_profile, target , outputs , dev , type, account, https_public_web, user, password, role, database, warehouse, schema}},
    keywordstyle={[2]\color{blue}},
    sensitive=true,
    literate=
      {:}{{{\color{blue}{:}}}}1
}

\usepackage{threeparttable}
\usepackage{balance}

\begin{document}
\title{ELT-Bench: An End-to-End Benchmark for Evaluating AI Agents on ELT Pipelines}

\author{Tengjun Jin}
\affiliation{%
  \institution{University of Illinois (UIUC)}
  \city{Urbana}
  \country{USA}
}
\email{tengjun2@illinois.edu}

\author{Yuxuan Zhu}
\affiliation{%
  \institution{University of Illinois (UIUC)}
  \city{Urbana}
  \country{USA}
}
\email{yxx404@illinois.edu}

\author{Daniel Kang}
\affiliation{%
  \institution{University of Illinois (UIUC)}
  \city{Urbana}
  \country{USA}
}
\email{ddkang@illinois.edu}

\begin{abstract}
Practitioners are increasingly turning to Extract-Load-Transform (ELT) pipelines with the widespread adoption of cloud data warehouses.
However, designing these pipelines often involves significant manual work to ensure correctness. Recent advances in AI-based methods, which have shown strong capabilities in data tasks, such as text-to-SQL, present an opportunity to alleviate manual efforts in developing ELT pipelines.
Unfortunately, current benchmarks in data engineering only evaluate isolated tasks, such as using data tools and writing data transformation queries, leaving a significant gap in evaluating AI agents for generating end-to-end ELT pipelines.

To fill this gap, we introduce ELT-Bench, an end-to-end benchmark designed to assess the capabilities of AI agents to build ELT pipelines. ELT-Bench consists of 100 pipelines, including 835 source tables and 203 data models across various domains. By simulating realistic scenarios involving the integration of diverse data sources and the use of popular data tools, ELT-Bench evaluates AI agents' abilities in handling complex data engineering workflows.
AI agents must interact with databases and data tools, write code and SQL queries, and orchestrate every pipeline stage. We evaluate two representative code agent frameworks, Spider-Agent and SWE-Agent, using six popular Large Language Models (LLMs) on ELT-Bench. The highest-performing agent, Spider-Agent Claude-3.7-Sonnet with extended thinking, correctly generates only 3.9\% of data models, with an average cost of \$4.30 and 89.3 steps per pipeline. Our experimental results demonstrate the challenges of ELT-Bench and highlight the need for a more advanced AI agent to reduce manual effort in ELT workflows. Our code and data are available at \url{https://github.com/uiuc-kang-lab/ELT-Bench}.

\end{abstract}
\maketitle

\pagestyle{\vldbpagestyle}
\begingroup

\section{Introduction}
Data engineers are increasingly leveraging Extract-Load-Transform (ELT) pipelines to integrate data and efficiently transform it into the required format as scalable cloud data warehouses become more accessible and storage prices continue to fall \cite{seenivasan2022etl, mbata2024surveypipelinetoolsdata, FOIDL2024111855, singhal2022etl, ETLhistory}. 
For example, the TPC-DI benchmark requires the creation of a decision support system for a retail brokerage firm by transforming data from various sources, including a trading system, internal Human Resources (HR), and Customer Relationship Management (CRM) systems \cite{TPCDI}. These data sources vary in formats, data types, attributes, and inter-table relationships \cite{TPCDI}. To build such a decision support system, data engineers can design an ELT pipeline: first, extracting and loading data into the data warehouse, followed by writing transformation queries to process the data for analysis.

Compared to traditional Extract-Transform-Load (ETL) pipelines, ELT pipelines ingest data directly into data warehouses, enabling real-time Business Intelligence (BI) analysis \cite{OnDemandELT}. Furthermore, with cloud infrastructure, ELT enhances scalability for processing large volumes of data \cite{ETLhistory} and offers greater flexibility in incorporating additional data transformations \cite{ETLvsELT}. These benefits make ELT pipelines an increasingly preferred choice for processing data across various scenarios \cite{seenivasan2022etl, mbata2024surveypipelinetoolsdata, FOIDL2024111855, singhal2022etl, ETLhistory}.

\begin{figure*}[t!]
    \centering
    \includegraphics[width=1\linewidth]{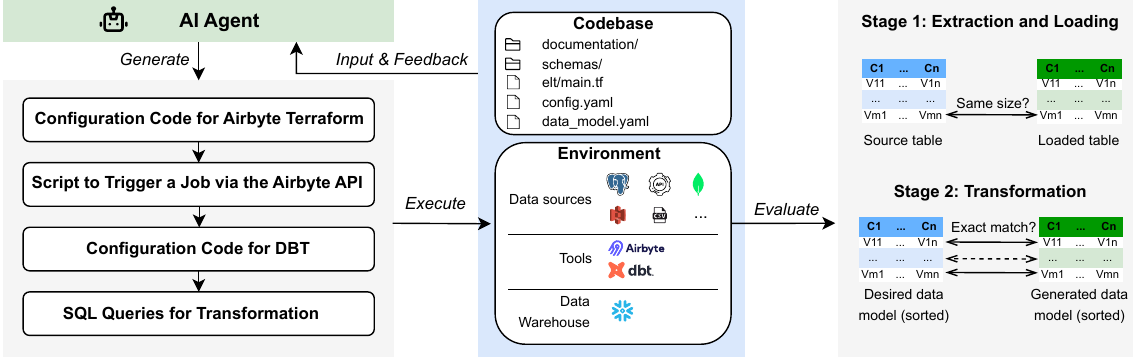}
    \caption{ELT-Bench is the first end-to-end benchmark designed to evaluate the ability of AI agents to build ELT pipelines. The agent must construct complete ELT pipelines from scratch by decomposing the complex workflow, interacting with databases and data tools, writing code and SQL queries, and calling APIs.}
    \label{fig:eltbench}
\end{figure*}

Developing ELT pipelines is an essential task for data engineers \cite{seenivasan2022etl, singhal2022etl, mbata2024surveypipelinetoolsdata, FOIDL2024111855, ETLhistory}, but the process requires significant manual work. Prior studies estimate that data engineers spend over 60\% of their time on data warehousing projects building data pipelines
\cite{effortestimation, datawarehouse, Researchdatawarehouse, realworlddata, Datamanipulation}. First, these pipelines must extract and integrate data from disparate sources with varying formats and standards. Second, data engineers or analysts need a deep understanding of the source data schema to write transformation queries.

Can AI agents effectively reduce the manual effort involved in constructing ELT pipelines? 
Recent advancements in Large Language Models (LLMs) have demonstrated strong capabilities in the text-to-SQL task, a crucial component of ELT pipelines. Notably, state-of-the-art (SOTA) techniques based on LLMs have achieved execution accuracy rates of 77.1\% and 91.2\% on the \bird \cite{bird} and Spider 1.0 \cite{spider1} benchmarks, respectively.
Researchers have recently developed AI agents to tackle more complex real-world tasks that demand reasoning, tool usage, planning, and memorization \cite{ReAct, Reflexion, sweagent, spider2, codeact, weng2023agent}. 
To evaluate the capability of emerging AI agents, researchers have proposed numerous benchmarks in the data domain \cite{ds1000, dacode, InfiAgent, spider2v, spider2}. However, there is no end-to-end benchmark designed with end-to-end ELT pipelines.

Building an end-to-end ELT benchmark is challenging because it requires sophisticated setup and configuration, time-consuming ground truth labeling, and thorough workflow verification to ensure reproducibility and correctness. 
First, annotators must set up various data management systems and platforms to store source data and provide data tools capable of handling diverse data sources. Second, annotators must prepare all necessary input files within the project base. Third, annotators must label ground truth by developing configuration files and writing complex transformation queries involving various relational operations (e.g., casting, type conversion, joins, aggregation, and ranking).
Finally, annotators manually execute and verify each ELT pipeline to validate the correctness of both the configured environment and generated annotations.



We address these challenges by introducing ELT-Bench, a new benchmark of 100 ELT pipelines associated with 835 source tables and 203 data models across various domains. For a single ELT pipeline, we spend approximately 3 to 5 hours of manual effort setting up the environment, annotating input files and the ground truth, and building the entire pipeline for verification. Notably, 60\% of the pipelines require extracting and integrating data from five distinct categories of sources (APIs, cloud services, relational databases, NoSQL databases, and flat files). In addition, the ground truth for each pipeline, on average, involves 187 lines of code per configuration file and 200 SQL tokens (tokenized by whitespace \cite{spider2}) per data model.

ELT-Bench is the first benchmark that covers the entire workflow for building ELT pipelines, providing a comprehensive evaluation through several interconnected subtasks. As shown in Figure \ref{fig:eltbench}, ELT-Bench requires agents to construct an end-to-end ELT pipeline from scratch, encompassing two primary stages: (1) data extraction \& loading stage and (2) data transformation stage. Agents execute their actions within a sandbox environment, which includes preinstalled packages and an established project base. This setup replicates the real-world workflow of a data engineer, challenging AI agents to break down the complex workflow into manageable subtasks, interact with databases and data tools, write relevant code, orchestrate every stage of the ELT pipeline, and finally generate the required data models.

We evaluate two code agent frameworks, Spider-Agent \cite{spider2} and SWE-Agent \cite{sweagent}, with six popular LLMs on ELT-Bench. The top-performing agent, Spider-Agent Claude-3.7-Sonnet with extended thinking, achieves a success rate of 57\% in the data extraction \& loading stage but only a success rate of 3.9\% in the data transformation stage. On average, Spider-Agent Claude-3.7-Sonnet consumes \$4.30 and requires 89.3 execution steps per task. Current agents' poor performance and high costs highlight the need for further advancements in AI agents to reduce manual effort in developing ELT pipelines.

\section{ELT-Bench}
In this section, we first introduce the data source of ELT-Bench, followed by summary statistics of tasks. We then provide an overview of ELT-Bench and outline the annotation pipeline. Finally, we demonstrate how an agent can complete one specific task in ELT-Bench as an example.

\subsection{Data Collection}
To ensure the quality of data, we collect databases based on a widely used text-to-SQL benchmark, \bird \cite{bird}, and the GitHub repository of an enterprise software, \texttt{Fivetran} \cite{Fivetran}.
\begin{itemize}[leftmargin=*]
    \item \bird is a text-to-SQL benchmark with large-scale databases spanning 37 domains. We use all the databases that have enough natural language questions to extract features as columns in data models, leading to 78 out of 80 open-source databases. Previous study indicates that databases in \bird can contain noise levels as high as 49\% \cite{birdnoise}. To ensure quality, we manually verify every natural language question and its corresponding SQL query used in our benchmark, correcting all identified errors.
    \item \texttt{Fivetran} is a data movement platform that develops \texttt{dbt} packages to facilitate the analysis of data from popular sources, such as Microsoft Advertising, Instagram Business, and YouTube Analytics. We sampled 22 databases from \texttt{Fivetran}.
\end{itemize}


\begin{table*}[h]
    \centering
    \caption{Comparison of ELT-Bench with two existing benchmarks in the data engineering field. ELT-Bench is the first end-to-end benchmark that focuses on building ELT pipelines, whereas Spider2-V concentrates on data tool usage and Spider 2.0 on the text-to-SQL workflow. Note: A single task may involve multiple data models and use both \texttt{Airbyte} and \texttt{DBT}, so ELT-Bench encompasses 203 data transformations and 200 data tools.}
    \label{tab:dataset_comparison}
    \begin{tabular}{l c c c c c}
    \toprule
    \textbf{Benchmark} & \textbf{\# Tasks} & \textbf{Data Extraction \& Loading} & \textbf{Data Transformation} & \textbf{Data Tools} & \textbf{End-to-End} \\
    \midrule
    Spider2-V \cite{spider2v} & 494 & \LeftCentered{\mycheck~(48)} & \LeftCenteredx{\mycross} & \LeftCenteredxx{\mycheck~(410)} & \LeftCenteredxxx{\mycross} \\
    Spider 2.0 \cite{spider2} & 632 & \LeftCentered{\mycross}      & \LeftCenteredx{\mycheck~(120)} & \LeftCenteredxx{\mycheck~(68)} & \LeftCenteredxxx{\mycross} \\
    \midrule
    ELT-Bench                 & 100 & \LeftCentered{\mycheck~(100)} & \LeftCenteredx{\mycheck~(203)} & \LeftCenteredxx{\mycheck~(200)} & \LeftCenteredxxx{\mycheck~(100)} \\
    \bottomrule
    \end{tabular}

\end{table*}

\subsection{Benchmark Statistics}
ELT-Bench contains 100 ELT pipelines associated with 835 source tables and 203 data models. As shown in Table \ref{tab:dataset_comparison}, compared to existing agent benchmarks for data engineering, ELT-Bench is the first end-to-end benchmark that covers the entire ELT pipeline construction workflow. 
In contrast, Spider 2-V \cite{spider2v} focused on evaluating an agent's ability to use high-level data tools individually, such as using \texttt{Airbyte} to extract and load data and using \texttt{DBT} with a given SQL query to transform data.
It does not include writing low-level SQL queries for data transformation or creating a complete pipeline using multiple tools. Furthermore, Spider 2.0 \cite{spider2} focuses on general text-to-SQL workflows, with only 10.8\% of tasks involving \texttt{DBT}. 
We highlight the characteristics of ELT-Bench as follows.

\begin{table}[t]
\caption{Statistics of ELT-Bench, illustrating the distribution of data sources, source tables, lines of \texttt{Terraform} code, data models, and SQL tokens per data model. As shown, ELT-Bench consists of ELT pipelines that involve multiple data sources, extensive code, and complex data transformations.}
\label{tab:elt_statistics}
    \centering
    \begin{tabular}{ll}
        \toprule
        \textbf{Statistics} & \textbf{Number}\\ 
        \midrule
        \textbf{\# Categories of Data Sources} & 100  \\ 
        \hspace{1em} 2 data sources & 7  \\ 
        \hspace{1em} 3 data sources & 15 \\ 
        \hspace{1em} 4 data sources & 18 \\ 
        \hspace{1em} 5 data sources & 60 \\ 
        \midrule
        \textbf{\# Source Tables} & 100  \\ 
        \hspace{1em} < 5 tables & 36  \\ 
        \hspace{1em} 5 - 10 tables & 40 \\ 
        \hspace{1em} > 10 tables & 24 \\  
        \midrule
        \textbf{\# Lines of Airbyte Terraform Code} & 100  \\ 
        \hspace{1em} < 100 lines & 7  \\ 
        \hspace{1em} 100 – 200 lines & 63 \\ 
        \hspace{1em} > 200 lines & 30 \\ 
        \midrule
        \textbf{\# Data Models} & 100  \\
        \hspace{1em} 1 data model & 50  \\ 
        \hspace{1em} 2 data models & 22  \\ 
        \hspace{1em} $\geq$ 3 data models & 28  \\
        \midrule
    \makecell[l]{\textbf{\# SQL Tokens per Data Model} \\ (Tokenized by whitespace \cite{spider2})} & 100  \\
        \hspace{1em} < 100 tokens & 8  \\ 
        \hspace{1em} 100–200 tokens & 19  \\ 
        \hspace{1em} > 200 tokens & 73  \\
        \bottomrule
    \end{tabular}
\end{table}

\minihead{Diverse Data sources} 
As shown in Tables \ref{tab:elt_statistics} and \ref{tab:data_sources}, our benchmark features diverse data sources. In total, 60 pipelines require extracting data from 5 categories of data sources, and 24 pipelines involve extracting more than 10 tables. Furthermore, 30 pipelines require writing more than 200 lines of code in \texttt{Terraform} files to extract data from these sources and load them into the data warehouse.

\minihead{Complex Data Transformation}
ELT-Bench evaluates the agent’s ability to write SQL queries based on natural language to generate target data models. Specifically, 28 pipelines require generating at least three data models. Following the approach in Spider 2.0 \cite{spider2}, we tokenize the SQL queries using whitespace and then count the resulting tokens to measure complexity. Because pipelines from \texttt{Fivetran} include both a staging and an intermediate layer, we calculate the average number of tokens per data model for each pipeline. As shown in Table \ref{tab:elt_statistics}, 73 pipelines demand over 200 tokens per data model, illustrating the complexity of these SQL queries.

\subsection{ELT-Bench Overview}
\minihead{Task Description}
ELT-Bench requires agents building an end-to-end ELT pipeline from an existing project base, which contains connection information, target data models, an initialization file for data tools, schemas of source tables, and data tool documentation. The pipeline must extract data from a variety of sources, load it into a data warehouse, and finally transform the source data into the target data models.

\minihead{Model Inputs}
We now describe the details of the pre-established project base, which consists of the following files:
\begin{enumerate}[leftmargin=*]
\item \texttt{config.yaml} contains the necessary connection information for data sources, data warehouses, \texttt{Airbyte}, and \texttt{DBT}. For example, extracting data from \texttt{PostgreSQL} requires specifying the host, port, user, password, schema, database, and tables.

\item \texttt{data\_model.yaml} defines the data models the ELT pipeline generates. Each data model includes a description, column names, and explanations for each column.

\item \texttt{elt/main.tf} contains the code to initialize \texttt{Terraform} provided in \texttt{Airbyte}.

\item \texttt{documentation}: This directory includes extracted or modified documentation from \texttt{Airbyte}, providing guidance on writing configuration code and triggering sync jobs.

\item \textit{schemas}: This directory contains the column names and descriptions of source tables.
\end{enumerate}

\minihead{Enviroment Description} 
ELT-Bench also includes a complex environment, as it requires agents to interact with a variety of data storage platforms and data tools. We describe them in detail below:
\begin{enumerate}[leftmargin=*]
\item \textit{Data sources}: We select five common categories of data sources for ELT-Bench, as shown in Table \ref{tab:data_sources}. We use Docker containers to deploy four data sources, including \texttt{PostgreSQL}, \texttt{MongoDB}, \texttt{REST API}, and \texttt{Amazon S3} (simulated using \texttt{LocalStack} \cite{localstack}), while providing downloadable links for flat files.

\item \textit{Data warehouse}: We use \texttt{Snowflake} \cite{snowflake} as our data warehouse to store extracted data and execute transformation queries that generate target data models, as it is a widely studied and popular cloud data warehousing solution \cite{spider2, CloudBenchmark}.

\item \textit{Data tools}: We use \texttt{Airbyte} \cite{airbyte} for data integration, a leading open-source data integration tool for ELT pipelines, which has also been used in prior work \cite{spider2v}. \texttt{Airbyte} runs in a separate Docker container. To generate target data models, we use \texttt{DBT} \cite{dbt}, a widely adopted data transformation tool \cite{spider2, spider2v}.

\item \textit{Packages and functions}: We provide a Docker file with all the required packages. Additionally, since data extraction and loading jobs typically take several minutes, we provide a script that monitors the status of all synchronization jobs and waits for their completion. This prevents redundant \texttt{Airbyte} API and LLM calls, reducing execution steps and costs.
 
\end{enumerate}




\subsection{Annotation Pipeline}
We now describe the annotation pipeline of ELT-Bench, which consists of following six steps:

\begin{table*}[t!]
\caption{Overview of common data source categories, representative sources, and their real-world applications.}
    \label{tab:data_sources}
    \centering
    \renewcommand{\arraystretch}{1.2} 
    \begin{tabular}{l|l|p{7.5cm}}
        \toprule
        \textbf{Data Source Category} & \textbf{Representative Sources} & \textbf{Applications in Practice} \\ 
        \midrule
        APIs & \texttt{REST API} & Web services, third-party platforms, real-time applications. \\ 
        Cloud Services & \texttt{Amazon S3} & Big data platforms, modern applications. \\ 
        Relational Databases & \texttt{PostgreSQL} & Traditional enterprise systems, transactional systems. \\ 
        NoSQL Databases & \texttt{MongoDB} & Modern web applications, real-time data systems. \\ 
        Flat Files & CSV, JSONL, Parquet & Third-party data providers,  backups. \\ 
        \bottomrule
    \end{tabular}
\end{table*}

\minihead{Step 1: data sources convertion}
To simulate integrating various data sources in a real-world ELT pipeline, we convert the collected data into different formats based on its characteristics and the classifications in Table~\ref{tab:data_sources}. The original \bird data is stored in \texttt{Sqlite}, while \texttt{Fivetran} data is in CSV format. Data is typically stored in multiple formats in real-world scenarios, depending on its intended use. For example, as shown in Table \ref{tab:data_sources}, relational databases are commonly used for transactional data storage. The selection of a target format follows these criteria:
\begin{enumerate}[leftmargin=*]
    \item We identify the potential data sources based on Table~\ref{tab:data_sources}.
    \item  We select the format that maximizes the source diversity if a data source can be represented in multiple formats.
    \item We ensure that the selected format is compatible with \texttt{Airbyte}.
\end{enumerate}

\minihead{Step 2: data source and environment setup}
The second step involves setting up the environment for storing data in different formats. As mentioned, we use Docker containers to deploy various data sources. We write the necessary scripts for data storage, including table creation and data insertion for \texttt{PostgreSQL} and \texttt{MongoDB}, \texttt{REST API} implementation, and data upload scripts for cloud storage. Furthermore, because the existing \texttt{Airbyte} extractor does not support local APIs, we have developed a custom extractor that can retrieve data from a \texttt{REST API} running inside a Docker container. We only need to perform all of these setup steps once before running the experiments.

\minihead{Step 3: configuration annotation}
After converting data formats and setting up the environment, we annotate the necessary connection information for data storage platforms and tools, which serve as one of the inputs in ELT-Bench. Notably, the annotated configurations do not strictly match the field names in the \texttt{Airbyte} documentation. This setting increases complexity and requires the agent's reasoning capabilities. For example, when extracting data from \texttt{MongoDB}, we specify the connection string as follows:
\begin{lstlisting}[style=yamll,frame=single,columns=fixed,breaklines=true,keywordstyle=\color{black}]
mongodb:
  config:
    connection_string: 'mongodb://elt-mongodb:27017/?directConnection=true'
\end{lstlisting}
The agent must determine that \texttt{MongoDB} is self-managed and configure \texttt{Airbyte} based on the provided configuration:
\begin{lstlisting}[style=terraform, frame=single,  columns=fixed,breaklines=true]
configuration = {
  database_config = {
    self_managed_replica_set = {
      connection_string = "mongodb://eltmongodb:27017/directConnection=true"}}}
\end{lstlisting}

\minihead{Step 4: data model definition}
In this step, we annotate the columns and their corresponding descriptions of data models in each database. Each data model consists of descriptive attributes from dimension tables and quantitative metrics from fact tables, representing specific entities. We categorize the columns in the data model based on the transformations applied:

\begin{enumerate}[leftmargin=*]
    \item \textit{Derived columns} come from either direct copies of the columns in source tables or transformations through basic operations (e.g., renaming, concatenation, and mathematical operations). 
    
    
    \item \textit{Aggregated columns} summarize data from the fact table using aggregation functions such as \texttt{SUM}, \texttt{AVG}, \texttt{COUNT}, \texttt{MAX} and \texttt{MIN}. 
    
    
    \item \textit{Categorical columns} classify data into predefined categories based on thresholds or conditions. \\
    \textit{Example:} The \texttt{has\_more\_than\_10\_movies\_1960\_to\_1985} is set to 1 if the director has directed at least 10 movies between 1960 to 1985; otherwise set to 0.
    
    \begin{lstlisting}
SELECT director_id, CASE WHEN COUNT(*) > 10 THEN 1 
  ELSE 0 END AS has_more_than_10_movies_1960_to_1985
FROM movies
WHERE movie_release_year BETWEEN 1960 AND 1985
GROUP BY director_id;
    \end{lstlisting}
    
    \item \textit{Ranked Columns} apply ranking functions (e.g., \texttt{RANK()}) to establish an order based on specific criteria and extract a value at a particular rank position.
    
    \textit{Example:} The \texttt{highest\_average\_score\_film} column shows the film directed by a director with the highest average score, with ties broken by the ascending order of the movie title.
    \begin{lstlisting}[morekeywords={RANK, WITH, OVER, PARTITION}]
WITH rated_film_ranks AS (
  SELECT director_id, movie_title, 
    RANK() OVER (PARTITION BY director_id 
      ORDER BY AVG(rating_score) DESC, 
      movie_title ASC) AS rated_film_rank
  FROM movie_platform.movies AS T1
  JOIN movie_platform.ratings AS T2 
    ON T1.movie_id = T2.movie_id
  GROUP BY director_id, movie_title)
SELECT director_id, 
  movie_title AS highest_average_score_film
FROM rated_film_ranks
WHERE rated_film_rank = 1
    \end{lstlisting}
\end{enumerate}

We now describe how we extract these four types of columns for our data models based on the natural language questions of \textsc{Bird}. While the original questions are typically designed for a specific entity, such as a single movie or person, we extract features for all entities instead. For example, consider the question, "Which film directed by Abbas Kiarostami has the highest average score?" This corresponds to the extracted feature \texttt{highest\_average\_score\_film} in the \texttt{Directors} data model, which represents the film with the highest average score for each director. After extracting features, we add text descriptions for these columns to help the agent better understand the desired data model and reduce ambiguity.

To make the data transformation stage in ELT-Bench more challenging, we rank the original SQL queries by complexity and prioritize features that involve more SQL components, conditions, and table joins \cite{spider1}. Each data model typically consists of three derived columns from the source tables and five additional columns (i.e., aggregated columns, categorical columns, and ranked columns) extracted from \bird questions.
Since the databases from \texttt{Fivetran} already contain predefined data models, we directly refine these models by removing columns generated by utility functions and those that contain only null values, unless they are required by another data model (a data model may incorporate another data model as an intermediate stage).


\minihead{Step 5: ground-truth annotation}
We annotate the ground truth using the configuration file and the defined data models. First, we review the official documentation to understand each configuration field and implement the necessary code accordingly.
Next, we annotate the SQL queries used to generate the defined data models in Step 4. To achieve this, we initially validate the existing queries provided by the \bird benchmark and the \texttt{Fivetran} repository. We modify those valid queries to conform to our defined data models; otherwise, we write gold queries from scratch.


\minihead{Step 6: execution-based verification}
To ensure the quality and correctness of ELT-Bench, we manually execute and verify each ELT pipeline, thoroughly confirming the accuracy of environment configurations and annotations.

In the first stage, we validate whether \texttt{Airbyte} can correctly extract data from diverse sources and load it into the data warehouse based on the provided configurations. Since \texttt{Airbyte} is an actively developing project, we encounter several issues:  
\begin{enumerate}[leftmargin=*]
    \item \textit{Table name case sensitivity in \texttt{PostgreSQL}}: \texttt{Airbyte} automatically converts table names in \texttt{PostgreSQL} to lowercase, which can cause a table not found error if the provided configuration contains table names with uppercase letters. 
    \item \textit{Schema detection for API data sources}: \texttt{Airbyte}’s automatic schema detection fails when the source data exceeds the maximum allowed string length.
\end{enumerate}
To address these issues, we standardize table names in \texttt{PostgreSQL} to only use lowercase letters and manually define schemas for affected API data sources.

In the data transformation stage, we verify our annotated SQL transformation queries. Specifically, for each defined data model, we write ten additional representative testing queries on average and then execute them against both the original source tables and corresponding data models. These testing queries encompass: (1) aggregation queries, (2) limit queries, and (3) queries obtained from \bird corresponding to entity-specific questions.
For any discrepancies or mismatches observed during query execution, we carefully review the annotated transformation queries, correct identified errors, and revise the transformation queries accordingly.

\subsection{Task Example}
We describe the process of building an ELT pipeline to generate a \texttt{customers} data model extracted from the \texttt{retails} database to demonstrate the task. The pipeline involves all five types of data sources. Starting in a sandbox environment with all required packages, we divide the task into two stages:

\begin{enumerate}[leftmargin=*]
\item \textit{Data extraction \& loading stage:} We use \texttt{Airbyte} to extract data from all data sources and load it into \texttt{Snowflake}.  First, we initialize \texttt{Airbyte} and write \texttt{Terraform} codes to configure \texttt{Airbyte}, all data sources, \texttt{Snowflake}, and the connections between data sources and \texttt{Snowflake} using information from the \texttt{config.yaml} file (Figure \ref{sub:input_config}) and relevant documentation. We illustrate the code for configuring the flat file 
\texttt{nation.jsonl} and establishing its connection to \texttt{Snowflake} (Figure \ref{sub:ouput_config}). Next, we execute \texttt{terraform apply}, which will apply the configuration code and generate configuration information in an output file. Finally, we extract connection IDs from the generated output file, trigger synchronization jobs via the \texttt{Airbyte} API, and monitor their status until completion.


\item \textit{Data transformation stage:} If all synchronization jobs in the first stage complete successfully, we proceed to data transformation using \texttt{DBT}. First, based on the provided configuration \texttt{config.yaml} (Figure \ref{sub:input_config}), we initialize a \texttt{DBT} project configured for \texttt{Snowflake} and set all necessary parameters (Figure \ref{sub:ouput_dbt}). Next, by referring to the \texttt{customers} data model definitions specified in \texttt{data\_model.yaml} (Figure \ref{subfig:input_dm}), we develop the transformation query (Figure \ref{sub:ouput_dm}). After executing \texttt{dbt run} to generate the data model, we validate the output by running \texttt{SELECT * FROM customers} in \texttt{Snowflake} to identify any issues that \texttt{DBT} might have missed.
\end{enumerate}

\begin{figure}[t!]
  \begin{subfigure}[t]{\columnwidth}
    \begin{lstlisting}[style=yamll]
Airbyte:
  config:
    files_definition_id: <id_1>
    workspace_id: <id_2>
flat_files:
- format: jsonl
  path: "https://..."
  sync_mode: full_refresh_append
  table: nation
snowflake:
  config:
    account: <account_id>
    database: retails
    password: <snowflake_password>
    username: AIRBYTE_USER ...
    \end{lstlisting}
  \caption{A partial configuration of \texttt{Airbyte} for the \texttt{retails} database, as defined in the provided \texttt{config.yaml} file.}
  \label{sub:input_config}
  \end{subfigure}

  \vspace{1em} 
\begin{subfigure}[t]{\columnwidth}
    \begin{lstlisting}[style=yaml]
models:
- name: customers
  description: Each record represents a customer.
  columns:
  - name: c_custkey
    description: Unique identifier for the customer.
  - name: order_date_highest_total_price
    description: The order date with the highest total price the customer has made, with ties broken by the ascending order of the order date. ...
    \end{lstlisting}
    \caption{A partial description of the \texttt{customers} data model for the \texttt{retails} database, as defined in the provided data\_model.yaml file.}
    \label{subfig:input_dm}
\end{subfigure}
  \caption{An example of provided input files of the \texttt{retails} database in ELT-Bench.}
  \label{fig:input_retails}
\end{figure}

\begin{figure}[t!]
  \begin{subfigure}[t]{\columnwidth}
    \begin{lstlisting}[style=terraform]
resource "airbyte_source_file" "jsonl_file_nation" {
  name          = "JSONL File nation"
  definition_id = "<id_1>"
  workspace_id  = "<id_2>"
  configuration = {
    dataset_name = "nation"
    format       = "jsonl"
    provider = {https_public_web = {}}
    url = "https://..." }}
resource "airbyte_connection" "nation_to_snowflake" {
  name = "JSONL nation to Snowflake"
  configurations = {
    streams = [
      {name = "nation"
       sync_mode = "full_refresh_append"}]}} ...
\end{lstlisting}
\caption{Partial code of Airbyte Terraform to configure
the flat file data source and its connection to \texttt{Snowflake} for the \texttt{retails} database.}
\label{sub:ouput_config}
\end{subfigure}

 \vspace{1em}
\begin{subfigure}[t]{\columnwidth}
    \begin{lstlisting}[style=dbt]
my_dbt_profile:
  target: dev
  outputs:
    dev:
      type: snowflake
      account: <account_id>
      user: AIRBYTE_USER
      password: <snowflake_password>
      database: retails ...
    \end{lstlisting}
    \caption{Example \texttt{DBT} configuration for the \texttt{retails} database in \texttt{Snowflake}.}
    \label{sub:ouput_dbt}
\end{subfigure}

  \vspace{1em}
\begin{subfigure}[t]{\columnwidth}
    \begin{lstlisting}[frame=single, morekeywords={WITH, RANK}]
WITH order_date_highest_total_price AS (
  SELECT o_custkey, o_orderdate,
    RANK() over(PARTITION by o_custkey
      ORDER BY o_totalprice DESC, o_orderdate
    ) AS price_rank FROM retails.airbyte_schema.orders)
SELECT T1.c_custkey AS c_custkey,
T2.o_orderdate AS order_date_highest_total_price, ...
FROM retails.airbyte_schema.customer T1
LEFT JOIN order_date_highest_total_price T2 ON T1.c_custkey = T2.o_custkey
  AND T2.price_rank = 1 ...
    \end{lstlisting}
    \caption{Partial code of the SQL transformation query for generating the \texttt{customers} data model.}
    \label{sub:ouput_dm}
\end{subfigure}

\caption{Files required to build the ELT pipeline.}
\label{fig:output_retails}
\end{figure}

\section{Experiments}
We evaluate two representative code agent frameworks, Spider-Agent \cite{spider2} and SWE-Agent \cite{sweagent}, using six LLMs on ELT-Bench. In this section, we first introduce the evaluation metrics of ELT-Bench, followed by a detailed explanation of the experimental settings for both agents. Finally, we present the evaluation results.

\begin{table*}[h]
    \centering
    \caption{ELT-Bench evaluation results for all tested agents and LLMs. Spider-Agent Claude-3.7-Sonnet with extended thinking performs best, with a 57\% SRDEL and 3.9\% SRDT.}
    \label{tab:main_results}
\begin{tabular}{l c r r r r}
\toprule
\textbf{Agent Framework} & \textbf{LLM} & \textbf{SRDEL (\%)} & \textbf{SRDT (\%) } & \textbf{Average Cost (\$)} & \textbf{Average Steps} \\
\midrule
& Claude-3.5-Sonnet & 23\% & 0 & 3.51 & 63.3 \\
 & GPT-4o & 15\% & 0 & 2.03 & 43.7 \\
Spider-Agent & DeepSeek-R1 & 0 & 0 & 0.38 & 18.4 \\
& Llama-3.1-405B & 0 & 0 & 0.39 & 22.0 \\
 & Qwen2.5-Coder-32B-Instruct  & 0 & 0 & 0.50 & 37.3 \\
\midrule
& Claude-3.5-Sonnet & 37\% & 1\% & 5.22 & 60.0 \\
 & GPT-4o & 0 & 0 & 5.22 & 114.3 \\
SWE-Agent & DeepSeek-R1 & 0 & 0 & 3.16 & 66.9 \\
& Llama-3.1-405B & 0 & 0 & 2.90 & 73.9 \\
 & Qwen2.5-Coder-32B-Instruct  & 0 & 0 & 0.48 & 39.1 \\
\midrule
Spider-Agent & Claude-3.7-Sonnet w/ extended thinking & 57\% & 3.9\% & 4.30 & 89.3 \\

\bottomrule
\end{tabular}

\end{table*}

\subsection{Evaluation Metrics}
We use the widely adopted metric, success rate \cite{spider2, spider2v, MLAgentBench, WebArena}, to assess the performance of agents on ELT-Bench. To provide a more comprehensive evaluation, we measure the success rate for both the data extraction \& loading stage and the data transformation stage. Specifically, we introduce the \textbf{Success Rate for Data Extraction \& Loading (SRDEL)} to measure the proportion of ELT pipelines that successfully extract and load data in the first stage and the \textbf{Success Rate for Data Transformation (SRDT)} to measure the proportion of data models successfully built in the second stage. Additionally, we measure the agent’s \textbf{average cost} (calculated based on token usage and API pricing \cite{openaipricing, Anthropicpricing, fireworks}) and \textbf{average steps} per task to assess its efficiency. We describe SRDEL and SRDT below.


\minihead{SRDEL} We evaluate the metric SRDEL in the first stage:
\begin{align*}
\text{SRDEL} &= \frac{\text{\# successful pipelines in data extraction \& loading}}{\text{\# total pipelines}},
\end{align*}
which measures the proportion of pipelines that successfully extract and load data.

A pipeline is considered successful in the data extraction \& loading stage if the pipeline successfully extracts data from all sources and loads it into the data warehouse. To evaluate this, we execute the following query for each source table in the data warehouse:

\begin{lstlisting}
SELECT COUNT(*) FROM source_table;
\end{lstlisting}
We then verify whether the query result matches the corresponding row count in the original data.

\minihead{SRDT} To evaluate the performance of the agent in the second stage, we use the metric SRDT:
\begin{align*}
\text{SRDT} &= \frac{\text{\# correctly generated data models}}{\text{\# total data models}},
\end{align*}
which measures the proportion of correctly generated data models among all data models (one ELT pipeline may involve multiple data models). To assess the correctness of a generated data model, we execute the following query:
\begin{lstlisting}
SELECT * FROM data_model ORDER BY unique_key;
\end{lstlisting}
The unique key may consist of a composite set of columns determined manually for each data model to ensure the query produces consistent results across different runs. We use this query to create CSV files for the generated data model, which are then compared against the ground truth, which is also derived from the same query.

A generated data model is correct if it contains all columns of the ground truth. Following prior work \cite{spider2}, we permit extra columns in the generated data model since they do not affect functionality.




\subsection{AI Agent Frameworks}
We select two representative code agent frameworks: Spider-Agent \cite{spider2} and SWE-Agent \cite{sweagent} since ELT-Bench requires capabilities such as file viewing and editing, code generation, and command execution. As baseline evaluations, we combine these two agent frameworks with five LLMs, including GPT-4o \cite{gpt4ocard}, Claude-3.5-Sonnet \cite{TheC3}, two open-sourced LLMs (Llama-3.1-405B-Instruct \cite{llama3}, Qwen2.5-Coder-32B-Instruct \cite{qwen2025}), and one reasoning model (DeepSeek-R1 \cite{deepseekr1}). In addition, as a case study aimed at exploring the frontier reasoning model, we also evaluate Spider-Agent Claude-3.7-Sonnet with extended thinking on ELT-Bench.

\minihead{Spider-Agent}
Spider-Agent is a code agent framework designed for database-related tasks, providing command-line interfaces for multi-turn interactions with environments \cite{spider2}. It also enables direct interaction with databases to extract detailed source table information (e.g., column values) and verify the correctness of transformation queries (e.g., \texttt{DBT} may fail to detect format errors). 
The agent employs the ReAct \cite{ReAct} framework, in which the LLM generates thought and decides the next action based on current observation and history trajectory at each iteration. We use the default parameter settings of Spider-Agent, except for changing the maximum allowed steps to 100, as ELT-Bench presents more challenging tasks compared to Spider 2.0 \cite{spider2}.


\minihead{SWE-Agent} 
SWE-Agent is a code agent framework designed to address GitHub issues \cite{sweagent}. Compared to Spider-Agent, it does not allow direct database access. In each iteration, the agent interacts with the filesystem based on its observations. SWE-Agent operates as a function-calling agent by prompting the LLM to invoke predefined functions, and it also offers a thought-action mode for LLMs that lack native support for tool usage. Consequently, we employ function calling for GPT-4o and Claude-3.5-Sonnet, and the thought-action approach for Llama-3.1-405B-Instruct, Qwen2.5-Coder-32B-Instruct, and DeepSeek-R1. We apply the default parameter settings of SWE-Agent, with one modification: retaining the last 25 observations for the agent due to the complexity of ELT-Bench. Following prior work~\cite{sweagent}, we allocate a same cost budget to all evaluated LLMs. To establish a comparable budget for both SWE-Agent and Spider-Agent, we first estimate the cost of completing 100 agent steps using Spider-Agent across all LLMs. We then select the highest of these estimates and round it up to the nearest integer, yielding a budget of \$6 for each evaluated LLM.



\subsection{Evaluation Results}

We report our evaluation metrics for all evaluated agents and LLMs in Table~\ref{tab:main_results}. The poor performance, high cost, and extensive step requirements highlight the challenges of ELT-Bench. The top-performing agent, Spider-Agent Claude-3.7-Sonnet with extended thinking, attains a 57\% success rate for data extraction \& loading, but only a 3.9\% success rate for data transformation. Despite these limitations, this agent demonstrates substantial performance improvements over the best-performing agent employing non-reasoning models, SWE-Agent Claude-3.5-Sonnet, with 54.1\% improvement in the data extraction and loading stage and 290\% improvement in the data transformation stage. 
Moreover, ELT-Bench presents a significantly higher computational cost compared to Spider 2.0 \cite{spider2}. While 30 agent steps are sufficient for most tasks in Spider 2.0, with an average cost of \$0.30 per instance using Spider-Agent GPT-4o, evaluating Spider-Agent GPT-4o on ELT-Bench requires an average of 43.7 steps and costs \$2.03 per task. 

We present a detailed error analysis for the baseline agent evaluations in Section~\ref{sec:error_analysis}, followed by an in-depth case study of Spider-Agent Claude-3.7-Sonnet in Section~\ref{sec:case_study}.



\begin{figure}
    \centering
    \includegraphics[width=1\linewidth]{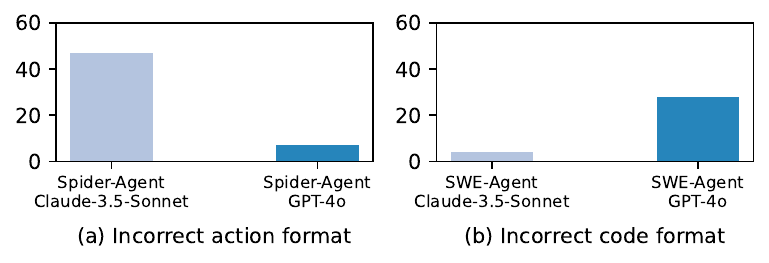}
    \caption{The number of tasks with incorrect formats. }
    \label{fig:inc_fmt}
\end{figure}

\begin{figure}[t]
  \begin{subfigure}[t]{\columnwidth}
    \begin{lstlisting}[style=terraform]
Action: CreateFile(filepath='/root/.dbt/profiles.yml':
```retail_complains: ...```<@\textcolor{red}{)}@>
\end{lstlisting}
\caption{Incorrect action format.}
\label{sub:wrong_fmt_sonnet}
\end{subfigure}

 \vspace{1em}

  \begin{subfigure}[t]{\columnwidth}
    \begin{lstlisting}[style=terraform]
provider "airbyte" {username = "<username>"}<@\textcolor{red}{\}}@>
\end{lstlisting}
\caption{Incorrect code format.}
\label{sub:wrong_fmt_gpt_4o}
\end{subfigure}

\caption{Incorrect action format generated by Spider-Agent \claude and incorrect code format generated by SWE-Agent GPT-4o.}
\label{fig:wrong_code_fmt}
\end{figure}

\section{Error Analysis}
\label{sec:error_analysis}
In this section, we examine the errors encountered by different agents and LLMs. We first highlight common issues observed among open-source LLMs. Then, we provide a detailed analysis of the errors arising in the data extraction \& loading stage and the data transformation stage for Spider-Agent and SWE-Agent using \claude and GPT-4o. 

\subsection{Error Analysis of Open-Sourced LLMs}
In ELT-Bench, the environment starts with a project base requiring agents to use official documentation extracted from the \texttt{Airbyte} \texttt{Terraform} website \cite{airbyteterraform}, reflecting a realistic scenario where data engineers must learn from documentation to use data tools, especially for those underdeveloped tools. 
However, open-sourced LLMs struggle to interact with the provided project base, resulting in a 0\% success rate in the data extraction \& loading stage. 
In addition, the complexity of ELT-Bench necessitates that the agent maintain a substantial memory length, leading to excessive prompt length issues when employing Qwen2.5-Coder-32B-Instruct.

\minihead{Failure to check configuration information} We first analyzed the performance of Spider-Agent DeepSeek-R1 and found that it failed primarily by neglecting the provided \texttt{config.yaml}. Spider-Agent DeepSeek-R1 references \texttt{config.yaml} in only five tasks. Instead, in most cases, it generates the configuration with random values, which results in execution errors in \texttt{Airbyte}. Even in the five tasks where \texttt{config.yaml} is used, the agent either produces an incomplete configuration or inserts its generated actions directly into the file rather than executing the actions.

\minihead{Failure to consult documentation files}
We further analyzed the remaining agents' performances and frequently observed a failure to reference the provided documentation. Without consulting up-to-date documentation, agents instead generate configuration code based on outdated versions of \texttt{Airbyte}'s documentation, resulting in errors related to incorrect resource types, as exemplified by Action 6 in Figure~\ref{sub:incor_resource}.
Specifically, we tracked how frequently they consulted the \texttt{Snowflake} configuration guide. Despite explicit prompt instructions, neither Spider-Agent nor SWE-Agent, when using Qwen2.5-Coder-32B-Instruct, consult this documentation. In contrast, Spider-Agent Llama-3.1-405B references the guide three times, SWE-Agent Llama-3.1-405B references it 20 times, and SWE-Agent DeepSeek-R1 refers to it 22 times. Consequently, SWE-Agent Llama-3.1-405B successfully configures the \texttt{Snowflake} destination in 9 of the 20 instances. Similarly, SWE-Agent DeepSeek-R1 correctly configures this destination in 10 of the 22 instances. However, these agents encountered other issues such as incorrect action formatting and failures in triggering synchronization jobs.


\minihead{Excessive prompt length}
Another common issue for agents using Qwen2.5-Coder-32B-Instruct is the excessive prompt length error \cite{spider2}, caused by exceeding its maximum supported context length of 32,767 tokens. Specifically, we observe this issue occurring in 15 tasks with Spider-Agent Qwen2.5-Coder-32B-Instruct and 87 tasks with SWE-Agent Qwen2.5-Coder-32B-Instruct. This discrepancy arises because Spider-Agent retains only the last 25 steps of model outputs and environment observations, whereas SWE-Agent retains all model outputs starting from the first step, along with only the last 25 environment observations.

\begin{figure}
    \centering
    \includegraphics[width=1\linewidth]{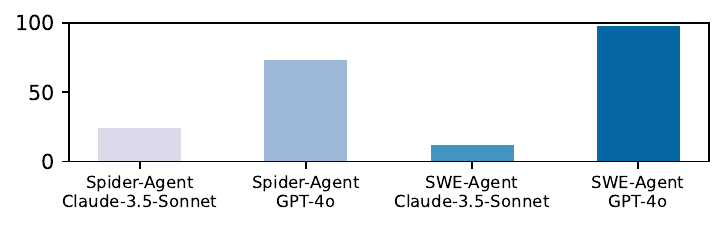}
    \caption{The number of tasks with incorrect \texttt{Snowflake} password field. }
    \label{fig:inc_sf}
\end{figure}

\begin{figure}
    \centering
    \includegraphics[width=1\linewidth]{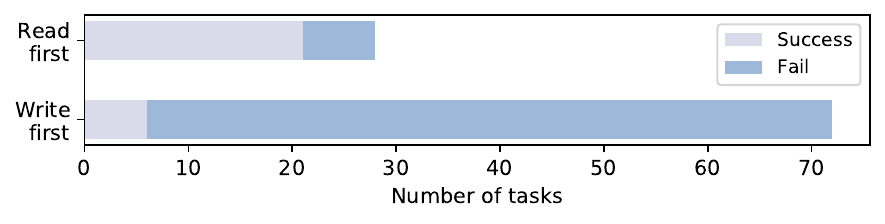}
    \caption{The success and failure rates of Spider-Agent GPT-4o in Stage 1 under two strategies: when reading the documentation first (27\% of tasks), it achieves a 78.8\% success rate; when writing the configuration first (73\% of tasks), the success rate drops to 9.6\%.}
    \label{fig:action_order}
\end{figure}

\begin{figure}[t]
  \begin{subfigure}[t]{\columnwidth}
   \input{figures/example/action_step}
\caption{The execution path of a successful task in Stage 1. The agent writes the configuration after reading the documentation.}
\label{sub:cor_resource}
\end{subfigure}

 \vspace{1em}

  \begin{subfigure}[t]{\columnwidth}
    \input{figures/example/failed_steps}
\caption{The execution path of a failed task in Stage 1. The agent writes the configuration file before reading the documentation and only fixes the detected error after reading the documentation.}
\label{sub:incor_resource}
\end{subfigure}

\caption{Comprasion the execution path of a successful task and a failed task.}
\label{fig:exe_path}
\end{figure}

\subsection{Error Analysis of Data Extraction \& Loading}
\label{subsec:stg1_error}
We examined the common issues encountered by Spider-Agent and SWE-Agent when using GPT-4o and Claude-3.5-Sonnet, including incorrect action or code formats, incorrect \texttt{Snowflake} password fields, incorrect table sizes, and missing configuration for multiple flat files. We describe each of these problems in detail below.

\minihead{Failure to generate action or code in the required format}
We observed that LLMs frequently generate actions in incorrect formats when used with Spider-Agent, and similarly produce code in incorrect formats under SWE-Agent.
Spider-Agent, which builds on the ReAct framework \cite{ReAct}, mandates that the LLM produce a valid action at each iteration, terminating the process if three formatting errors are detected via regular expressions. Despite providing detailed descriptions and examples of all actions, LLMs can still generate invalid actions. For instance, in Figure~\ref{sub:wrong_fmt_sonnet}, Spider-Agent \claude incorrectly places a closing parenthesis on the last line instead of before the colon in the first line, causing a parsing failure. As shown in Figure ~\ref{fig:inc_fmt}, Spider-Agent \claude terminates 47\% of tasks in Stage 1 because of unparsable actions, while Spider-Agent GPT-4o only terminates 7\% of tasks. 

In contrast, SWE-Agent employs a function-calling framework with well-defined functions, which avoids parsing errors. However, misformatted code can still be generated. For instance, SWE-Agent GPT-4o generates an extra right curly brace at the end of a code block, as illustrated in Figure~\ref{fig:wrong_code_fmt}. We show in Figure ~\ref{fig:inc_fmt} that SWE-Agent \claude produces misformatted code in 4\% of cases, whereas SWE-Agent GPT-4o exhibits a 28\% error rate.
These findings highlight the importance of developing frameworks that robustly ensure LLMs generate correct syntax for actions and code.

\begin{figure}
    \centering
    \includegraphics[width=1\linewidth]{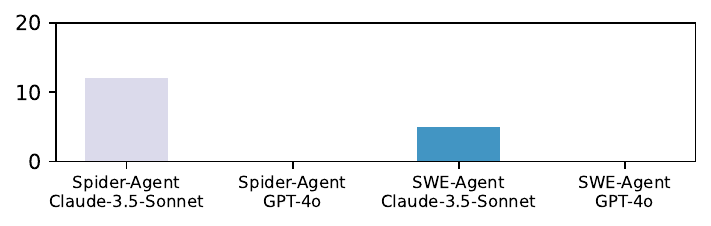}
    \caption{The number of tasks with incorrect table size. }
    \label{fig:inc_size}
\end{figure}

\begin{figure}
    \centering
    \includegraphics[width=1\linewidth]{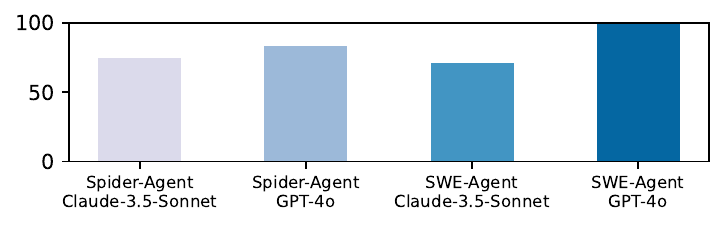}
    \caption{The failure rate of configuring multiple flat files. The total number of tasks with multiple flat files is 24. }
    \label{fig:mul_file}
\end{figure}

\begin{figure}[t!]
  \begin{subfigure}[t]{\columnwidth}
  \centering
\includegraphics[width=\linewidth]{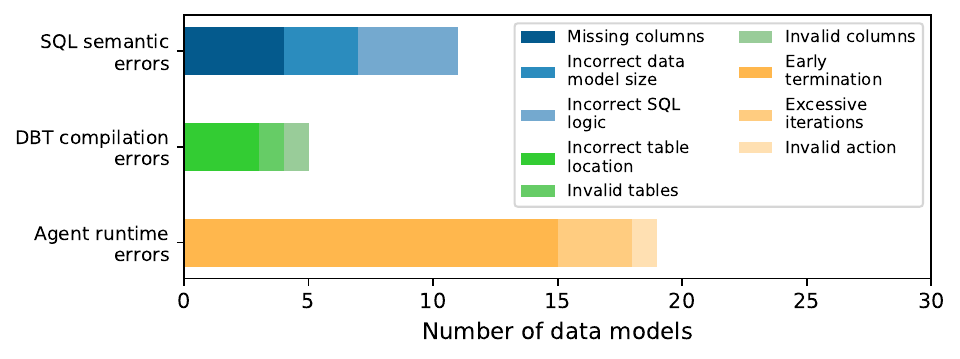}
\caption{Statistics of Spider-Agent GPT-4o in the second stage.}
\label{sub:stg_spider_4o}
\end{subfigure}

\vspace{1em}
  \begin{subfigure}[t]{\columnwidth}
  \centering
\includegraphics[width=\linewidth]{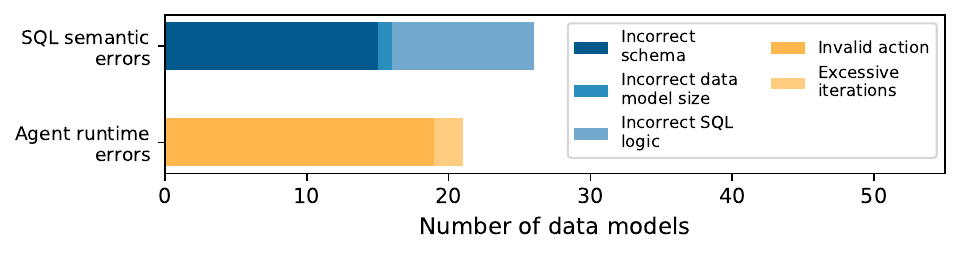}
\caption{Statistics of Spider-Agent \claude in the second stage.}
\label{sub:stg_spider_sonnet}
\end{subfigure}

\vspace{1em}
  \begin{subfigure}[t]{\columnwidth}
  \centering
\includegraphics[width=\linewidth]{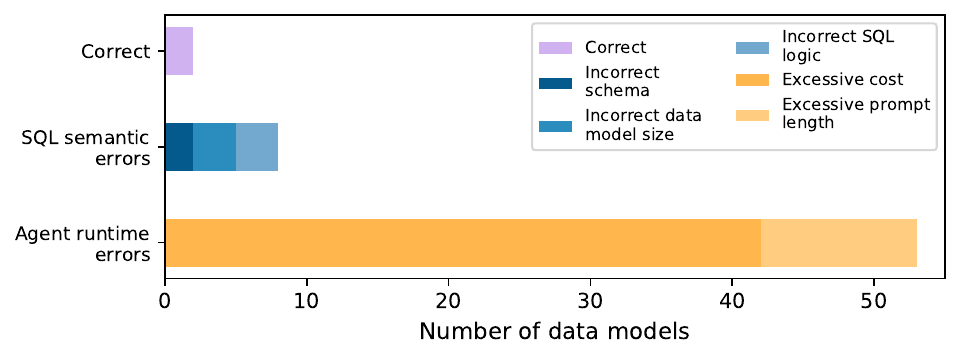}
\caption{Statistics of SWE-Agent \claude in the second stage.}
\label{sub:stg_swe_sonnet}
\end{subfigure}

\caption{Statistics of agent performance on generating data models in the second stage. Each subfigure includes results for databases that the agent successfully completed in the first stage (35, 47, and 63 data models, respectively).}
\label{fig:stage2}
\end{figure}

\minihead{Failure to configure the \texttt{Snowflake} password field}
We examined the \texttt{Snowflake} password field, which must be written in the format as shown in Figure~\ref{sub:cor_resource}. However, as Figure \ref{fig:inc_sf} illustrates, SWE-Agent GPT-4o fails in up to 98\% of tasks to configure the \texttt{Snowflake} password field. In contrast, SWE-Agent \claude performs much better, failing to configure it in only 12\% of tasks. 

We further analyzed the execution path of Spider-Agent GPT-4o and identified two distinct strategies the agent adopted when configuring \texttt{Airbyte} \texttt{Terraform}. In one strategy (Figure~\ref{sub:cor_resource}), the agent attempts to write the configuration code first and then runs \texttt{terraform apply -auto-approve}. Upon encountering an error indicating an incorrect resource type, the agent consults the documentation but only corrects the specific issue reported by \texttt{Terraform}. Because \texttt{Airbyte} \texttt{Terraform} ignores any fields that are not explicitly defined, other misconfigurations remain undetected, which finally causes the ELT pipeline to fail.

In contrast, when the agent references the documentation before writing the configuration, it is more likely to produce a valid \texttt{Terraform} configuration, leading to a higher success rate for data extraction \& loading. As illustrated in Figure~\ref{fig:action_order}, the agent reads the documentation before writing the configuration in 27 tasks and successfully configures the \texttt{Snowflake} password field in 21 tasks. By comparison, in 73 tasks, the agent writes the configuration first, and only six tasks succeed. These observations underscore the importance of the agent's effective planning (e.g., executing actions in the correct sequence) in achieving higher success rates.

\minihead{Incorrect table size due to repeated synchronization job triggers}
We observed that, in some cases, the size of the source tables did not match the size of the original data. Analyzing the execution paths of failed cases, we found that the agent repeatedly triggered the same synchronization job. For example, if the original dataset contains 100 rows but the agent executes the synchronization job three times, the resulting table in \texttt{Snowflake} ends up with 300 rows instead of the intended 100. As shown in Figure~\ref{fig:inc_size}, Spider-Agent \claude repeatedly triggers the same synchronization job in 12 tasks.
These findings highlight the importance of short-term memorization in the agent for tracking executed actions and preventing redundant synchronization jobs.

\minihead{Missing configuration for multiple flat files}
For \texttt{Postgres}, \texttt{MongoDB}, APIs, and \texttt{Amazon} \texttt{S3}, multiple tables or files can be configured within a single source block and a single connection block.
In contrast,
\texttt{Airbyte} \texttt{Terraform} requires individual source and connection configuration blocks for each flat file. ELT-Bench includes 24 instances to evaluate whether the agent can correctly generate multiple configuration blocks for multiple flat files. As shown in Figure~\ref{fig:mul_file}, SWE-Agent GPT-4o fails on all 24 instances, while even the best-performing agent, SWE-Agent Claude-3.5-Sonnet, still fails in 70.8\% of cases. These findings emphasize the need for the agent to handle diverse configuration patterns across different data sources.

\subsection{Error Analysis of Data Transformation}
\label{subsec:stg2_error}
We evaluated the data transformation stage performance of Spider-Agent GPT-4o, Spider-Agent Claude-3.5-Sonnet, and SWE-Agent Claude-3.5-Sonnet, as these agents completed the data extraction \& loading stage for some databases. Specifically, Spider-Agent GPT-4o successfully processes 15 databases in Stage 1, comprising 35 data models; Spider-Agent \claude processes 23 databases, comprising 47 data models, while SWE-Agent \claude processes 37 databases, comprising 63 data models.

However, only SWE-Agent \claude successfully generates two correct data models. We categorize Stage 2 errors into three main types: agent runtime errors, \texttt{DBT} compilation errors, and SQL semantic errors, and show the performance in Figure~\ref{fig:stage2}.

\minihead{Agent runtime errors}
Agent runtime errors primarily result from configuration or action-related issues that cause the agent to terminate before generating a data model. For example, due to a \$6 maximum cost constraint, SWE-Agent \claude fails to create 42 data models within the allotted budget. In addition, Spider-Agent GPT-4o fails to generate 15 data models because it prematurely halts in Stage 1.
Among these prematurely terminating cases, 46.7\% stop because, although the agent correctly assumes it should proceed to Stage 2, it mistakenly executes a terminate action instead. Meanwhile, 53.3\% fail due to incorrect synchronization job status checks. As shown in Figure~\ref{sub:stg_swe_sonnet}, agent runtime errors account for up to 84.1\% of failed data model generations.

\minihead{DBT compilation errors}
We observed \texttt{DBT} compilation errors in Spider-Agent GPT-4o, which were identified by \texttt{DBT} during \texttt{dbt run}. As shown in Figure ~\ref{sub:stg_spider_4o}, these errors occur because the agent fails to correctly specify the corresponding database and schema for tables used in the query (3 data models) or references non-existent tables (1 data model) or columns (1 data model).

\minihead{SQL semantic errors}
SQL semantic errors occur when the agent successfully creates a data model in the database, but the model is incorrect. We classify the semantic errors of data models in ascending severity: incorrect schema, missing columns, incorrect data model size, and flawed SQL logic. For example, if a data model is placed in the wrong schema and omits some columns, we categorize it as an incorrect schema error. In Spider-Agent GPT-4o (Figure~\ref{sub:stg_spider_4o}), the most common SQL semantic errors (11 data models) are due to missing columns (4 data models) and flawed SQL logic (4 data models). For Spider-Agent Claude-3.5-Sonnet, 26 data models contain SQL semantic errors, with incorrect schema assignments (15 data models) and flawed SQL logic (10 data models), as shown in Figure~\ref{sub:stg_spider_sonnet}. Similarly, as illustrated in Figure~\ref{sub:stg_swe_sonnet}, SWE-Agent Claude-3.5-Sonnet produces 8 data models with SQL semantic issues, including incorrect schema (2 data models), incorrect data model size (3 data models), and flawed SQL logic (3 data models).

\section{Case Study}
\label{sec:case_study}

\begin{figure}
    \centering
    \includegraphics[width=1\linewidth]{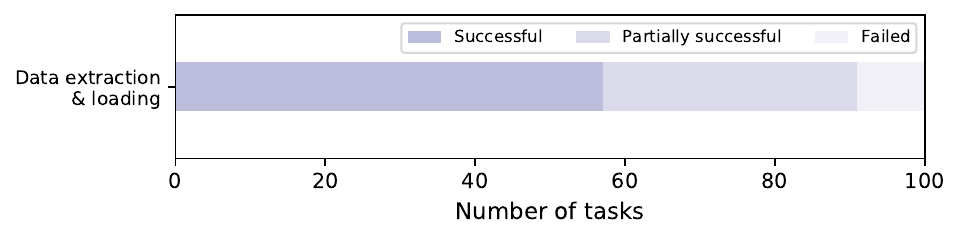}
    \caption{Task completion status in the data extraction \& loading stage. Spider-Agent Claude-3.7-Sonnet fails on all sources in nine tasks and partially succeeds in 34 tasks.}
    \label{fig:case_study}
\end{figure}
\begin{figure}
    \centering
    \includegraphics[width=1\linewidth]{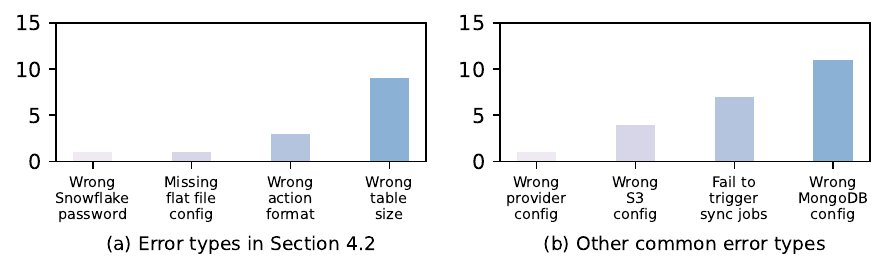}
    \caption{Common error types encountered by Spider-Agent Claude-3.7-Sonnet in the first stage.}
    \label{fig:case_study2}
\end{figure}

\begin{figure}
    \centering
    \includegraphics[width=1\linewidth]{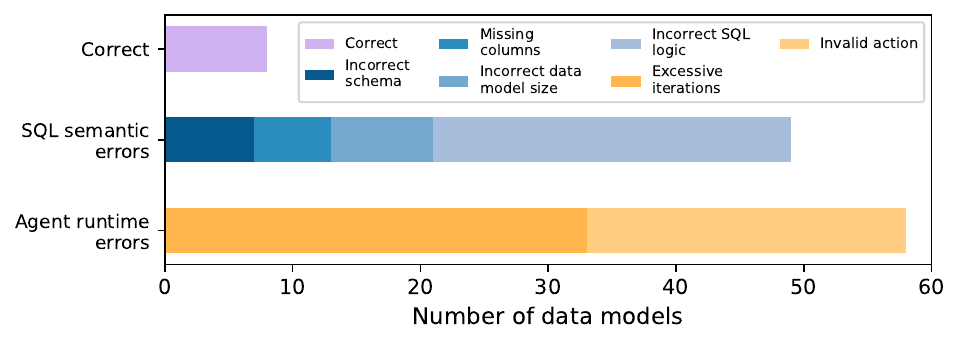}
    \caption{Statistics of Spider-Agent Claude-3.7-Sonnet in the second stage.}
    \label{fig:case_study3}
\end{figure}

\begin{figure}
    \centering
    \includegraphics[width=1\linewidth]{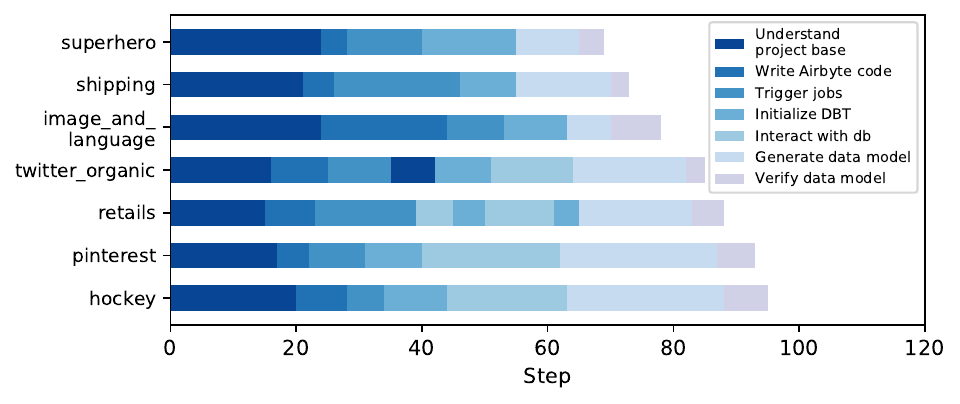}
    \caption{The action trajectories of the agent on databases with at least one successful data model.}
    \label{fig:case_study4}
\end{figure}

In this section, we present an in-depth analysis of the Spider-Agent Claude-3.7-Sonnet with extended thinking, focusing on its performance and the errors encountered across two stages of the task. We then examine its action trajectories in successful cases.

Spider-Agent Claude-3.7-Sonnet achieves a significant performance improvement of 54.1\% over the second-best agent, SWE-Agent Claude-3.5-Sonnet, in the first stage. As shown in Figure~\ref{fig:case_study}, Spider-Agent Claude-3.7-Sonnet fails on all data sources in the first stage for nine tasks. We further analyzed common error types during the first stage, with results depicted in Figure~\ref{fig:case_study2}. Our initial examination of the four issue types described in Section~\ref{subsec:stg1_error} reveals that Spider-Agent Claude-3.7-Sonnet significantly reduced error frequencies across all categories compared to Spider-Agent Claude-3.5-Sonnet, achieving up to a 95.8\% error reduction. Further analysis of additional common issues, shown in the right part of Figure~\ref{fig:case_study2}, indicates that Spider-Agent Claude-3.7-Sonnet frequently fails on specific data source types, particularly \texttt{MongoDB}. This finding aligns with the partial success of 34 tasks observed in Figure~\ref{fig:case_study}.

Spider-Agent Claude-3.7-Sonnet demonstrates a 290\% performance improvement in the second stage compared to SWE-Agent Claude-3.5-Sonnet. As illustrated in Figure~\ref{fig:case_study3}, the primary issues of Spider-Agent Claude-3.7-Sonnet in the second stage include excessive iterations (28.7\%), incorrect SQL logic (24.3\%), and invalid actions (21.7\%). 

To better understand Spider-Agent Claude-3.7-Sonnet’s workflow, we illustrate the action paths of the agent for databases that successfully produced at least one correct data model in Figure~\ref{fig:case_study4}. On average, the agent executed 83.6 steps for each successful case. To provide clarity, we categorize these actions into defined phases based on the agent's thoughts and actions. Specifically, if fewer than five consecutive steps belonging to one phase appear between two occurrences of another identical phase, we group these intermediate steps into the surrounding phase. For instance, it is common for the agent to briefly interact with the database during the ``generate data model'' phase. As depicted in Figure~\ref{fig:case_study4}, the Spider-Agent Claude-3.7-Sonnet spends most of its execution steps to the phases of ``understanding the project base'' (averaging 20.6 steps) and ``generating the data model'' (averaging 17 steps).


\section{Sensitivity and Ablation study}
\subsection{Multiple Runs Improve Agent Performance}
We evaluated Spider-Agent GPT-4o’s performance on ELT-Bench with one attempt (pass@1) and five attempts (pass@5). As shown in Figure~\ref{fig:abl_pass}, Spider-Agent GPT-4o achieves a pass@5 rate of 57\% in Stage 1, indicating that in 57\% of tasks, at least one of the five attempts successfully extracts data from multiple sources and loads it into the data warehouse. This result represents a 3.8$\times$ improvement over its pass@1 performance. However, in Stage 2, despite having more successfully loaded source tables, Spider-Agent GPT-4o still fails to build a correct data model.

We further use the pass\textasciicircum k metric \cite{taubench} to evaluate the consistency and robustness of Spider-Agent GPT-4o on ELT-Bench. As shown in Figure~\ref{fig:pass_hat}, as the number of trials increases, pass\textasciicircum k for Spider-Agent GPT-4o drops significantly, eventually reaching 0 when k equals 5, indicating the need for a more robust agent in future work.

\subsection{Using Documentation Improves Agent Performance}
We evaluated whether Spider-Agent \claude and Spider-Agent GPT-4o could complete the data extraction \& loading stage without consulting documentation. Since LLMs are trained on a fixed knowledge cutoff, their ability to reference up-to-date documentation is crucial for completing real-world tasks. To assess their adaptability, we compared their performance in data extraction \& loading with and without documentation guidance.

In our experiments, we provided the agents with documentation on configuring \texttt{Airbyte} \texttt{Terraform} and invoking the \texttt{Airbyte} API to initiate synchronization jobs. As shown in Figure~\ref{fig:abl_doc}, Spider-Agent \claude and Spider-Agent GPT-4o exhibit degraded performance in the data extraction \& loading stage when documentation is unavailable. Without access to documentation, Spider-Agent \claude succeeds in only one task, while Spider-Agent GPT-4o fails in all tasks. These findings reveal that both \claude and GPT-4o rely not only on memorized knowledge but also on their reasoning abilities to complete tasks.


\begin{figure}
    \centering
    \includegraphics[width=1\linewidth]{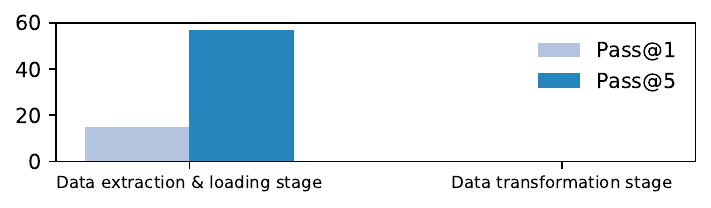}
    \caption{
The success rate of Spider-Agent GPT-4o with one versus five attempts. The success rate improves from 15\% to 57\% in the first stage but remains 0\% in the second stage.}
    \label{fig:abl_pass}
\end{figure}

\begin{figure}
    \centering
    \includegraphics[width=1\linewidth]{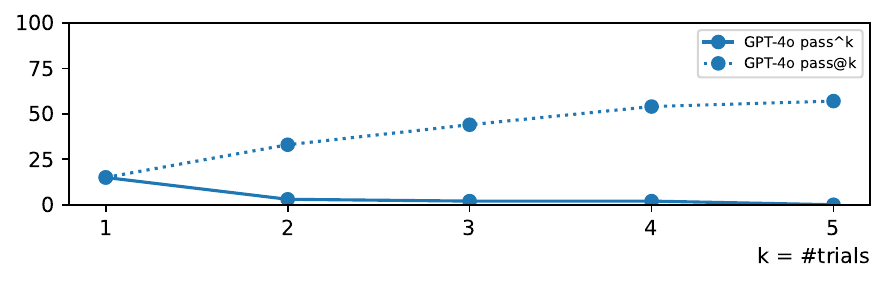}
    \caption{
Pass\textasciicircum k and pass@k in the first stage of ELT-Bench.}
    \label{fig:pass_hat}
\end{figure}

\begin{figure}
    \centering
    \includegraphics[width=1\linewidth]{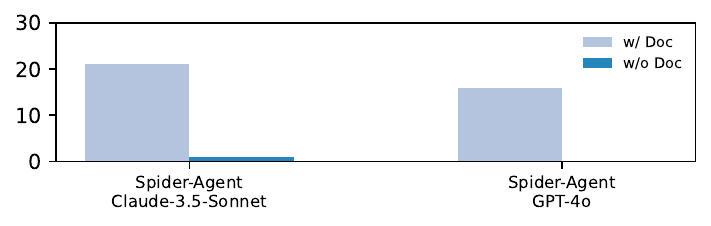}
    \caption{The success rate of Spider-Agent with Claude-3.5-Sonnet and GPT-4o in the data extraction \& loading stage, evaluated with and without documentation. Success rates decrease from 21\% to 1\% for Claude-3.5-Sonnet, and from 15\% to 0\% for GPT-4o.}
    \label{fig:abl_doc}
\end{figure}
\section{Related Work}
\minihead{ELT and ETL data pipelines}
ELT and ETL data pipelines are essential for converting raw data into structured, reliable formats, playing an important role in modern data engineering workflows. ETL techniques have been extensively studied over decades \cite{ETLhistory}, while the rise of cloud data warehousing has driven the increasing adoption of ELT pipelines \cite{seenivasan2022etl, mbata2024surveypipelinetoolsdata, FOIDL2024111855, singhal2022etl}. Early research mainly focus on conceptual modeling for ETL processes \cite{conceptualETL, DataMapping, UMLETL}. More recent efforts have aimed at automating various stages of ETL and ELT pipelines to minimize engineering effort, including Semantic Web-based approaches for attribute mapping \cite{webETL}, template-driven automatic data loading \cite{automatingloading}, and machine learning-based data integration \cite{MLETL}. In this work, we introduce ELT-Bench, a benchmark designed to facilitate the development of AI agents capable of automating ELT pipeline construction, thus reducing manual effort.

\minihead{Text-to-SQL benchmarks and methods}
Researchers have studied the text-to-SQL task, which aims to convert natural language queries into SQL queries, for decades. Early text-to-SQL datasets primarily target single database scenarios \cite{iyer2017learningneuralsemanticparser, SQLizer, Finegan_Dollak_2018}. More recent datasets, including WikiSQL~\cite{wikisql} and Spider~\cite{spider1}, extend this scope by introducing cross-domain scenarios requiring models to generalize to unseen databases.
The \bird benchmark is further introduced to evaluate text-to-SQL methods within large-scale database contexts, focusing on both query accuracy and execution efficiency~\cite{bird}. Initially, text-to-SQL methods primarily leverage graph neural networks (GNNs) \cite{cao2021lgesqllinegraphenhanced} and long short-term memory (LSTM) networks\cite{SQLNet}. Recent research has increasingly adopted fine-tuning techniques~\cite{XiYan-SQL, CodeS} and prompting approaches~\cite{dong2023c3zeroshottexttosqlchatgpt, gaotexttosql, pourreza2023dinsqldecomposedincontextlearning} to further enhance SQL generation accuracy with the advent of LLMs. ELT-Bench tasks agents with generating complex SQL transformation queries to construct data models based on provided column names and descriptions. These queries typically involve intricate structures, including 
nested subqueries and multi-table joins.


\minihead{AI agent benchmarks} 
To support the development of AI agents for solving complex real-world tasks, researchers have introduced diverse benchmarks across several domains, including software engineering \cite{swebench}, machine learning \cite{MLAgentBench}, and web-based interactions \cite{WebArena, WorkArena}. In the data domain, existing benchmarks primarily focus on data science code generation \cite{ds1000, dacode} and data analysis \cite{InfiAgent}. Additionally, Spider 2-V~\cite{spider2v} evaluates agents' proficiency in using data tools, while Spider 2.0~\cite{spider2} assesses agent performance on enterprise-focused text-to-SQL workflows. In contrast, ELT-Bench is the first benchmark designed to assess AI agents' capabilities in developing real-world, end-to-end ELT pipelines.

\minihead{AI agents}
LLM-based Agents have emerged as a promising approach for addressing real-world challenges across various fields, including software engineering \cite{sweagent, codeact, zhang2024autocoderoverautonomousprogramimprovement}, web browsing \cite{pan2024autonomousevaluationrefinementdigital,lai2024autowebglmlargelanguagemodelbased} and data science and engineering \cite{InfiAgent, hong2024datainterpreterllmagent, spider2}. These agents typically consist of four crucial modules: reasoning \cite{cot, kojima2023largelanguagemodelszeroshot, yao2023treethoughtsdeliberateproblem}, tool usage \cite{schick2023toolformerlanguagemodelsteach, qin2023toolllmfacilitatinglargelanguage}, planning \cite{shinn2023reflexionlanguageagentsverbal, yao2023reactsynergizingreasoningacting}, and memorization \cite{zhu2023ghostminecraftgenerallycapable}. In this work, we evaluate two code generation agent frameworks (SWE-Agent \cite{sweagent} and Spider-Agent \cite{spider2}) on ELT-Bench to assess their performance in constructing ELT pipelines.
\section{Conclusion}
We introduce ELT-Bench, a comprehensive end-to-end benchmark specifically designed for real-world ELT pipeline tasks in the data engineering domain. ELT-Bench aims to replicate realistic scenarios by providing environments for diverse data sources and integrating widely adopted data tools. The benchmark presents a substantial challenge, as the top-performing agent, Spider-Agent Claude-3.7-Sonnet, correctly generates data models in only 3.9\% of cases. This performance gap highlights significant opportunities for future research to develop more powerful and intelligent AI agents capable of handling complex ELT workflows.
\clearpage

\balance
\bibliographystyle{ACM-Reference-Format}
\bibliography{main}

\end{document}